\documentclass[11pt]{article}
\pdfoutput=1 

\usepackage{jheppub} 

\usepackage[T1]{fontenc} 
\usepackage{array}

\newcommand\fft[2]{\frac{#1}{#2}}
\newcommand\ft[2]{{\textstyle\frac{#1}{#2}}}
\newcommand\nn{\nonumber}

\title{\boldmath One-loop supergravity on $\mathrm{AdS}_4\times S^7/\mathbb{Z}_k$ and comparison with ABJM theory}

\author{James T. Liu}
\author{and Wenli Zhao}

\affiliation{Michigan Center for Theoretical
Physics, Randall Laboratory of Physics,\\
The University of Michigan, Ann Arbor, MI 48109-1040}

\emailAdd{jimliu@umich.edu}
\emailAdd{wlzhao@umich.edu}

\abstract{The large-$N$ limit of ABJM theory is holographically dual to M-theory on 
AdS$_4\times S^7/\mathbb{Z}_k$.  The 3-sphere partition function has been obtained via localization, and its leading behavior $F_{\text{ABJM}}^{(0)}\sim k^{1/2}N^{3/2}$ is exactly reproduced in the dual theory by tree-level supergravity.  We extend this comparison to the sub-leading $\mathcal O(N^0)$ order by computing the one-loop supergravity free energy as a function of $k$ and comparing it with the ABJM result.  Curiously, we find that the expressions do not match, with $F_{\text{SUGRA}}^{(1)}\sim k^6$, while $F_{\text{ABJM}}^{(1)}\sim k^2$.  This suggests that the low-energy approximation $Z_{\text{M-theory}}=Z_{\text{SUGRA}}$ breaks down at one-loop order.}

\preprint{MCTP-16-21}

\begin{document} 
\maketitle
\flushbottom

\section{Introduction}
\label{sec:intro}

The AdS/CFT correspondence is a remarkable duality between large-$N$ field theories and gravity in the bulk.  As such, it has passed many non-trivial tests at the leading order in the large-$N$ expansion.  One prominent example is the computation of the holographic Weyl anomaly \cite{Henningson:1998gx}, which for IIB string theory on AdS$_5\times X_5$ yields
\begin{equation}
c=a=\fft{N^2}4\fft{\pi^3}{\mathrm{vol}(X_5)},
\end{equation}
at tree-level in the supergravity limit.  This result has been extended to the $\mathcal{O}(1)$ level by performing a one-loop computation, where the states running in the loop come from the Kaluza-Klein spectrum on $X_5$
\cite{Bilal:1999ph,Bilal:1999ty,Mansfield:2000zw,Mansfield:2002pa,Mansfield:2003gs,Liu:2010gz,Ardehali:2013gra,Ardehali:2013xya,Ardehali:2013xla,Beccaria:2014xda}.  An interesting feature of the one-loop contribution to the holographic Weyl anomaly is that it only receives contributions from the shortened multiplets in the Kaluza-Klein tower.  As such, this provides a connection between the holographic central charges and the superconformal index \cite{Ardehali:2014zba,Ardehali:2014esa}.

While the Weyl anomaly is a feature of even-dimensional field theories, similar holographic computations have been performed for odd-dimensional theories.  One approach has been to focus on the holographic entanglement entropy which can be defined in arbitrary dimensions \cite{Myers:2010tj}.  Alternatively, the 3-sphere free energy $F$ has been conjectured to play the role of the $a$-anomaly in odd-dimensional CFTs \cite{Giombi:2014xxa}.  In this paper, we extend the one-loop tests of AdS/CFT to the odd-dimensional case by examining the $\mathcal O(1)$ contributions to $F$.  In particular, we compute the holographic one-loop ABJM sphere partition function in the M-theory limit and compare with the matrix model result.

The ABJM model is a three-dimensional $\mathcal{N}=6$ superconformal Chern-Simons-Matter (CSM) theory with gauge group $U(N)_{k}\times U(N)_{-k}$ \cite{Aharony:2008ug}. It is conjectured to be the holographic dual of IIA string theory on AdS$_4\times CP_3$ in the `t~Hooft limit with $\lambda\equiv N/k$ finite and the dual of M-theory on AdS$_4\times S^7/\mathbb Z_k$ in the limit $N\to \infty$ with $k^5\ll N$.  As an odd-dimensional CFT, it has vanishing Weyl anomaly.  However, the sphere partition function has been computed from the matrix model, and takes the form \cite{Marino:2011eh}: 
\begin{equation}
Z_{\text{ABJM}}=C^{-\frac{1}{3}}e^{A(k)} \text{Ai}\left[C^{-\frac{1}{3}}\left(N-\frac{1}{3k}-\frac{k}{24}\right)\right]+Z_{\text{Non-Perturbative}},
\label{eq:ZABJM}
\end{equation}
where $C={2}/{\pi^2 k}$.  Here $A(k)$ encodes certain quantum corrections, and can be computed in the IIA (i.e.\ planar) limit as the all-genus sum of the constant map contributions to the free-energy \cite{Hanada:2012si}:
\begin{equation}
    A(k)=-\fft{\zeta(3)}{8\pi^2}k^2+\fft16\log\fft{4\pi}{k}+2\zeta'(-1)
    -\fft13\int_0^\infty\fft{dx}{e^{kx}-1}\left(\fft3{x^3}-\fft1x-\fft3{x\sinh^2x}\right).
    \label{eq:Ak}
\end{equation}
It is furthermore conjectured that this expression remains valid in the M-theory limit that we are mostly interested in \cite{Hanada:2012si}.  In particular, when expanded for small $k$, it reproduces the perturbative series computed with the Fermi gas approach in \cite{Marino:2011eh}.

The ABJM free energy can be expanded in the large-$N$ limit with the result%
\footnote{Here we use the convention $F=-\log Z$.}
\begin{equation}
F_{\text{ABJM}}=\fft{\pi\sqrt2}3k^{1/2}N^{3/2}-\fft{\pi}{\sqrt{2k}}\left(\fft{k^2}{24}+\fft13\right)N^{1/2}+F_{\text{ABJM}}^{(1)}+\mathcal O(N^{-1/2}),
\label{eq:ABJMexpand}
\end{equation}
where
\begin{equation}
F_{\text{ABJM}}^{(1)}=\frac{1}{4}\log N-\frac{1}{4}\log k+\frac{5}{4}\log 2-A(k).
\label{eq:ABJM1}
\end{equation}
The holographic ABJM free energy was computed in \cite{Drukker:2010nc}, and is given at leading order in the M-theory limit by
\begin{equation}
F_{\text{SUGRA}}^{(0)}=\fft{\pi\sqrt2}3k^{1/2}N^{3/2}.
\end{equation}
This precisely matches the leading term in the expansion of the matrix partition function (\ref{eq:ABJMexpand}).
The $\mathcal O(N^{1/2})$ term does not follow from a standard loop expansion of supergravity, which would be given in powers of the 11-dimensional Newton constant, $G_{11}\sim N^{-3/2}$.  Instead, it arises as a quantum correction in M-theory, and in particular from a shifted relation between ABJM and M-theory parameters resulting from the eight-derivative $C_3R^4$ term \cite{Bergman:2009zh,Aharony:2009fc,Drukker:2011zy,Bhattacharyya:2012ye}, as anticipated in \cite{McLoughlin:2008he}.

Our present focus is on the $\mathcal O(1)$ contribution, $F_{\text{ABJM}}^{(1)}$, which is dual to the one-loop free-energy in M-theory.  The $\log N$ term in (\ref{eq:ABJM1}) has been identified as a universal contribution independent of the specific compactification, and is given by the zero modes of the heat kernel of the $\mathcal{N}=6$ supergravity on AdS$_4\times X_7$, \cite{Bhattacharyya:2012ye}. It is likely that this term is fully captured by the zero modes in the supergravity limit, as the heat-kernel expansion in odd dimensions (corresponding to the M-theory limit) does not yield a log term apart from the zero modes.  Moreover, contributions beyond the supergravity limit are not expected to affect the zero mode counting, as this ought to be a robust feature of the low energy (and hence supergravity) limit.

Although the non-zero modes do not contribute to the $\log N$ term in (\ref{eq:ABJM1}), they are nevertheless expected to contribute to the one-loop holographic free energy.  Therefore a natural question arises as to whether a one-loop supergravity calculation can fully reproduce the $\mathcal{O}(1)$ term given in (\ref{eq:ABJM1}).  We will perform this computation in the M-theory limit, where the dual of ABJM theory in low energy limit is given by 11-dimensional supergravity on AdS$_4\times S^7/\mathbb{Z}_k$.

On the ABJM theory side, the AdS/CFT dictionary at leading order gives the relation%
\footnote{This leading order relation is sufficient, as the anomalous radius shift responsible for the $\mathcal O(N^{1/2})$ term \cite{Bergman:2009zh} has no effect on the $\mathcal O(1)$ term.}
\begin{equation}\label{dictionary}
N=\frac{2}{k \pi^2}\left(\fft{L}{l_p}\right)^6,
\end{equation}
where $L$ is the AdS$_4$ radius and $l_p$ is the 11 dimensional Planck length. Under (\ref{dictionary}), the $\mathcal{O}(1)$ term, (\ref{eq:ABJM1}), then becomes 
\begin{equation}\label{fabjm}
F^{(1)}_{\text{ABJM}}=\frac{3}{2}\log \frac{L}{l_p}-\frac{1}{2}\log\fft{k\pi}8-A(k).
\end{equation}
On the supergravity side, we regulate the one-loop determinants by working with a $4+7$ dimensional split.  We use spectral zeta function methods for determinants in AdS$_4$ before summing over the Kaluza-Klein spectrum on $S^7/\mathbb{Z}_k$.  The one-loop free energy is then given schematically by
\begin{equation}\label{fsugra}
F^{(1)}_{\text{SUGRA}}\sim \zeta'(0)\sim (\zeta(0)+c_0)\log\Lambda L+ a(k),
\end{equation}
where $\Lambda$ is the volume cutoff in the one-loop determinants, $c_0$ is the zero mode contribution, $a(k)$ is a term only dependent on $k$, and both $\zeta'(0)$ and $\zeta(0)$ refer to the regulated quantities after summing over the Kaluza-Klein spectrum. Notice that both $F_{\text{SUGRA}}^{(1)}$ and $F_{\text{ABJM}}^{(1)}$ have undetermined constants, such as $l_p$ and $\Lambda$. Thus we would not expect to precisely match the two terms, unless through a judicial choice of these constants. Instead, we are interested in whether (\ref{fabjm}) and (\ref{fsugra}) have the same functional dependence on $k$. Specifically, we can look for whether they have the same asymptotic behavior for $k$ while remaining in the M theory limit by requiring $k^5 \ll N$.

As can be seen from (\ref{eq:Ak}), asymptotically $A(k)\sim k^2$, and thus we would predict similar behavior in $F_{\text{SUGRA}}^{(1)}$. Our calculation, however, shows that this is not the case. Asymptotically we find a leading $k^6$ behavior for $F_{\text{SUGRA}}^{(1)}$. A $k^2$ term is present in the asymptotic expansion, but the coefficient does not match with that in $A(k)$. 
While this may be viewed as a failure of AdS/CFT at the one-loop level, we instead suggest that what this indicates is that the supergravity computation is incomplete, and that additional M-theory contributions beyond the supergravity limit will ultimately lead to agreement between the holographic and field theoretic expressions.

Along these lines, it is worth emphasizing that the even-dimensional AdS calculation has distinct properties from the odd-dimensional case.  Apart from the vanishing of the holographic Weyl anomaly for even-dimensional AdS (odd-dimensional CFT), the isometry groups of $\text{AdS}_{2n}$ and $\text{AdS}_{2n+1}$ fall in different classes in the classification of semi-simple Lie algebras.  This has a direct consequence at one-loop level, as it changes the measure of the spectral zeta function (which is a Mellin transform of the heat kernel) and thus the functional form of the one-loop determinant \cite{hawking1977,PhysRevD.47.3339,Camporesi:1994ga}.  As a result, while the one-loop partition function vanishes in $\text{AdS}_{2n+1}$ when summed over the states of a long representation of supersymmetry, this no longer the case for $\text{AdS}_{2n}$.  Since massive string states fall into long representations, they do not affect the one-loop corrections to the holographic Weyl anomaly in AdS$_5$.  In contrast, however, massive states in M-theory can be expected to contribute to the AdS$_4$ free energy. For this reason, it is perhaps not so surprising that our one-loop supergravity result does not fully capture the $\mathcal O(1)$ contribution to the ABJM free energy.

The paper is organized as follows. In section~\ref{sec:KKspec}, we work out the Kaluza-Klein spectrum on the $S^7/\mathbb{Z}_k$ orbifold.  (The more technical aspects are presented in Appendix~\ref{app:specmodk}.) In section~\ref{sec:freeenergy}, we outline the one-loop computation using the spectral zeta function in $\text{AdS}_4$, and give the exact expression of $F_{\text{SUGRA}}^{(1)}$  as well as its asymptotic expansion in $k$. We comment on some of the subtleties associated with regularization of the KK sum in Appendix~\ref{app:reg}, while some lengthy expressions are presented in Appendix~\ref{app:c12}.  Finally, in section~\ref{sec:disc}, we comment on possible implications and some open questions of the functional disagreement between $F_{\text{ABJM}}^{(1)}$ and $F_{\text{SUGRA}}^{(1)}$.

\section{Kaluza-Klein spectrum on the $S^7/\mathbb{Z}_k$ orbifold}\label{sec:KKspec}

The $\mathcal O(1)$ contribution to the holographic free energy of ABJM theory on $S^3$ may be computed by evaluating the one-loop partition function of M-theory on AdS$_4\times S^7/\mathbb Z_k$.  We work in the supergravity limit and first reduce to AdS$_4$, so that we are left with evaluating one-loop determinants on global AdS$_4$.  In this section, we work out the Kaluza-Klein spectrum, and in the following section we compute the free energy by summing over the spectrum.

In order to describe the $\mathbb Z_k$ orbifold, we note that the transverse space to a stack of M2-branes can be identified (at least locally) with $\mathbb{C}^4$. The action of $\mathbb{Z}_k$ is then given by \cite{Aharony:2008ug} 
\begin{equation}\label{action}
z_i\to e^{\frac{2\pi i}{k}} z_i.
\end{equation}
This action does not have any fixed points and is in fact smooth for finite $k$. One can then consider the Hopf fibration map $p:\mathbb{C}^4\to \mathbb{CP}^3$ whose fiber when restricted to $S^7$ embedded in $\mathbb{C}^4$ is $S^1$. The quotient space is thus a lens space. The action in fact acts only on the fiber and it shrinks the radius of the circle. In the limit $k\to \infty$, the metric becomes degenerate, and the quotient becomes singular in that it truncates the principal $U(1)$ bundle to its base space. Such a truncation gives rise to IIA supergravity on AdS$_4\times \mathbb{CP}^3$ which is dual to the IIA limit of ABJM theory.

The Kaluza-Klein spectrum of the orbifold under the above action is given by branching the $\mathcal{N}=8$ KK spectrum, labeled by representations of $so(8)$, the Lie Algebra of the isometry group of $S^7$, into that of the $\mathcal{N}=6$ spectrum, labeled by representations of $su(4)\oplus u(1)$, corresponding to the isometry groups of $\mathbb{CP}^3$ and $S^1$. We then select the KK multiplets that are stable under the action (\ref{action}), i.e.\ the ones with $U(1)$ charge divisible by $k$. The zero sector of the branching problem of $so(8)\to su(4)\oplus u(1)$ is obtained in \cite{0264-9381-1-5-005}, and similarly, with 
\begin{equation}
8_{s}\to 6_0+1_{-2}+1_{2},
\end{equation}
we can obtain the full branching for the $\mathbb Z_k$ orbifold.  The KK spectrum branched into $su(4)\oplus u(1)$ is given in Table.~\ref{tbl:Zkspect},
where $E_0$ represents the lowest energy and $[a,b,c;d]$ represents
\begin{equation}
[a,b,c;d]=\sum_{r=0}^{N}(n-r+a,b,r+c)_{n-2r+d},
\end{equation}
where $(n-r+a,b,r+c)$ is the Dynkin label of $su(4)$ and $n-2r+d$ is the corresponding $U(1)$ charge. In each sum, $N$ should be determined by the highest entry of the Dynkin label of $so(8)$. For example, we have
\begin{equation}
(n-2,0,0,0)\to [-2,0,0;-2]=\sum_{r=0}^{n-2}(n-r-2,0,r)_{n-2r-2}.
\end{equation}
%

\begin{table}[t]
    \centering
    \setlength\extrarowheight{1pt}
      \begin{tabular}{ | l | l | p{8cm} | l |}
    \hline
    Spin & $so(8)$ & $su(4)\oplus u(1)$ & $E_0$ \\ \hline
    $2^{+}$ & $(n,0,0,0), n\geq 0 $  & $[0,0,0;0]$ & $\frac{n}{2}+3$ \\[2pt] \hline
    $\frac{3}{2}^{(1)}$ & $(n,0,0,1)$, $n\geq 0$ & $[0,1,0;0]+[0,0,0;-2]+[0,0,0;2]$ & $\frac{n}{2}+\frac{5}{2}$ \\[2pt] \hline
    $\frac{3}{2}^{(2)}$ & $(n-1,0,1,0),n\geq 1$ & $[0,0,0;-2]+[-1,1,-1;0]+[-1,0,1;0] $& $\frac{n}{2}+\frac{7}{2}$\\[2pt]    \hline
    $1^{-(1)}$& $(n,1,0,0),n\geq 0$& $[0,0,0;0]+[1,0,1;0]+[0,1,0;-2]+[0,1,0;2]$&$\frac{n}{2}+2$\\[2pt] \hline
    $1^{+}$&$(n-1,0,1,1),n\geq1$&$[0,0,0;0]+[-1,0,1;-2]+[0,1,0;-2]$\par${}+[-1,1,1;0]+[-1,1,-1;-2]+[-2,1,0;0]$\par${}+[-1,2,-1;0]+[0,0,0;-4]+[-1,0,1;2]$&$\frac{n}{2}+3$\\ \hline
    $1^{-(2)}$&$(n-2,1,0,0),n\geq2$&$[-2,0,0;-2]+[-1,0,1;-2]+[-2,1,0;-4]$\par${}+[-2,1,0;0]$&$\frac{n}{2}+4$\\ \hline
    $\frac{1}{2}^{(1)}$&$(n+1,0,1,0),n\geq 0$&$[2,0,0;0]+[1,1,-1;2]+[1,0,1;2]$&$\frac{n}{2}+\frac{3}{2}$\\[2pt] \hline
    $\frac{1}{2}^{(2)}$&$(n-1,1,1,0),n\geq 1$&$[0,0,0;-2]+[-1,1,-1;0]+[-1,0,1;0]$\par${}+[1,0,1;-2]+[0,0,2;0]+[0,1,0;0]$\par${}+[-1,1,1;-2]+[0,1,0;-4]+[-1,1,1;2]$\par${}+[-2,2,0;-4]+[-1,2,-1;2]$&$\frac{n}{2}+\frac{5}{2}$\\ \hline
    $\frac{1}{2}^{(2)}$&$(n-2,1,0,1),n\geq 2$&$[-1,0,1;0]+[-1,0,1;-4]+[-2,0,0;-4]$\par${}+[-2,0,0;0]+[-1,1,1;-2]+[-2,1,0;-2]$\par${}+[-2,1,0;-2]+[-2,1,0;2]+[-2,1,0;-6]$\par${}+[-2,2,0;0]+[-2,2,0;-4]$&$\frac{n}{2}+\frac{7}{2}$\\ \hline
    $\frac{1}{2}^{(4)}$&$(n-2,0,0,1),n\geq2$&$[-2,1,0;-2]+[-2,0,0;-4]+[-2,0,0;0]$&$\frac{n}{2}+\frac{9}{2}$\\[2pt] \hline
    $0^{+(1)}$&$(n+2,0,0,0),n\geq0$&$[2,0,0;2]$&$\frac{n}{2}+1$\\[2pt] \hline
    $0^{-(1)}$&$(n,0,2,0),n\geq0$&$[1,0,1;0]+[2,0,0;-2]+[0,0,2;2]$\par${}+[-1,2,-1;0]+[1,1,-1;0]+[-1,1,1;0]$&$\frac{n}{2}+2$ \\ \hline
    $0^{+(2)}$&$(n-2,2,0,0),n\geq2$&$[-1,1,1;0]+[-1,1,1;-4]+[-2,2,0;-6]$\par${}+[-2,-2,0;2]+[-1,0,1;-2]+[-2,2,0;-2]$\par${}+[-2,1,0;-4]+[-2,1,0;0]+[-2,0,0;-2]$\par${}+[0,0,2;-2]$&$\frac{n}{2}+3$\\ \hline
    $0^{-(2)}$&$(n-2,0,0,2),n\geq2$&$[-2,0,0;-2]+[-2,0,0;-6]+[-2,0,0;2]$\par${}+[-2,2,0;-2]+[-2,1,0;-4]+[-2,1,0;0]$&$\frac{n}{2}+4$\\ \hline
    $0^{+(3)}$&$(n-2,0,0,0),n\geq2$&$[-2,0,0;-2]$&$\frac{n}{2}+5$\\[2pt] \hline
\end{tabular}
    \caption{The Kaluza-Klein spectrum of 11-dimensional supergravity on AdS$_4\times S^7/\mathbb Z_k$.  The notation $[a,b,c;d]$ is explained in the text.}
    \label{tbl:Zkspect}
\end{table}

One can now select the multiplets that are divisible by $k$, and rewrite the full branching into the branching mod $k$, i.e.\ leaving only multiplets with $U(1)$ charge divisible by $k$. However, the result is somewhat lengthy, and is relegated to Appendix~\ref{app:specmodk}. Although it appears that the KK spectra for even $k$ and odd $k$ are different, they give rise to the same $F_{\textrm{ 1-loop}}$ as we shall see in the next section. That makes the even/odd behavior of the free energy of ABJM theory for finite $N$ observed in \cite{Okuyama:2011su} more intriguing, as one might expect the even/odd behavior of the free energy to be a quantum effect of M-theory and thus reflected in the KK spectrum. However, it does not seem to be the case. 

To make the $\mathcal{N}=6$ supersymmetry explicit, one can organize the KK spectrum in terms of the unitary irreducible representations of the supergroup $Osp(4|6)$. Such representations may be labeled by considering the bosonic subalgebra $so(2,3)\oplus so(6)$.  We then label the representation by $\mathcal{D}(E_0, j,h_1,h_2,h_3)$, where $E_0$ is the lowest energy (or conformal dimension from the CFT point of view), $j\in \frac{1}{2}\mathbb{N}^0$ is the spin, and $h_1, h_2,h_3$ are highest weights of $so(6)$, such that $h_1\geq h_2\geq|h_3|$. (Here we find it more convenient to use highest weight labels; they are related to Dynkin labels $(a,b,c)$ by $a=h_2-h_3$, $b=h_1-h_2$ and $c=h_2+h_3$.)
Unitarity is not guaranteed \emph{a priori} given arbitrary values of $(E_0, j,h_1,h_2,h_3)$, but gives rise to the following conditions \cite{Bhattacharya:2008zy,Dolan:2008vc}:
\begin{align}
E_0&>j+h_1+1,&&\mbox{long};\nn\\
E_0&=j+h_1+1,&&\mbox{regular short (semi-short)};\nn\\
E_0&=h_1,\quad j=0,&&\mbox{isolated short (BPS)}.
\end{align}

Examination of the KK spectrum on $S^7/\mathbb Z_k$ demonstrates that it consists only of isolated short representations, and can be classified as either $\frac{1}{2}$-BPS or $\frac{1}{3}$-BPS states of $\mathcal N=6$.  The contents of these multiplets are described in Tables~\ref{tbl:12BPS} and \ref{tbl:13BPS}.
The full $\mathcal N=8$ spectrum, branched into $\mathcal N=6$ supermultiplets, is then given by
\begin{align}
\mathcal D_{\mathcal N=8}\left(\fft{n}2+1,0,n+2,0,0,0\right)=\,&
\mathcal D_{\text{$\fft12$-BPS}}\left(\frac{n}{2}+1,0,\fft{n}2+1,\fft{n}2+1,\fft{n}2+1\right)_{-n-2}\nn\\
&\oplus\mathcal D_{\text{$\fft12$-BPS}}\left(\frac{n}{2}+1,0,\fft{n}2+1,\fft{n}2+1,-\fft{n}2-1\right)_{n+2}\nn\\
&\oplus\sum_{i=0}^{n}\mathcal D_{\text{$\fft13$-BPS}}\left(\frac{n}{2}+1,0,\fft{n}2+1,\fft{n}2+1,\frac{n}{2}-i\right)_{-n+2i}.
\label{eq:susybranch}
\end{align}
The orbifold spectrum is then obtained by only considering supermultiplets with $U(1)$ charges $q\equiv0$ mod $k$, where $q$ is given by the subscripted quantities in (\ref{eq:susybranch}).

\begin{table}[t]
    \centering
     \begin{tabular}{ |c | c |  c | c | c |}
    \hline
    $E_0~\backslash ~j$& 0 & $\frac{1}{2}$& $1$ & $\frac{3}{2}$\\ \hline
    $h$& $[h,h,h]$& & &  \\ \hline
    $h+\frac{1}{2}$& &$[h,h,h-1]$& &\\ \hline
    $h+1$& $[h,h,h-2]$&&$[h,h-1,h-1]$& \\ \hline
    $h+\frac{3}{2}$& &$[h,h-1,h-2]$&&$[h-1,h-1,h-1]$ \\ \hline
    $h+2$& $[h,h-2,h-2]$&&$[h-1,h-1,h-2]$& \\ \hline
    $h+\frac{5}{2}$& &$[h-1,h-2,h-2]$&&\\ \hline
    $h+3$&$[h-2,h-2,h-2]$&&&\\ \hline
    
    \end{tabular}

    \caption{$\frac{1}{2}$-BPS multiplets $\mathcal D_{\text{$\fft12$-BPS}}(h,0,h,h,h)$ of $osp(4|6)$. Here $[h_1,h_2,h_3]$ are $so(6)$ highest weight labels.  The conjugate multiplet $\mathcal D_{\text{$\fft12$-BPS}}(h,0,h,h,-h)$ may be obtained by taking $h_3\to-h_3$.}
    \label{tbl:12BPS}
\end{table}

\begin{table}[t]
\centering
\begin{tabular}{|c|c|p{12cm}|}
\hline
$E_0$&$j$& $[h_1,h_2,h_3]$ \\ \hline
$h$&$0$&$[h,h,r]$ \\ \hline
$h+\fft12$&$\fft12$&$[h,h,r+1]\oplus[h,h,r-1]\oplus[h,h-1,r]$ \\ \hline
$h+1$&$0$& $[h,h,r]\oplus[h,h,r+2]\oplus[h,h,r-2]\oplus[h,h-1,r+1]\oplus[h,h-1,r-1]$\par${}\oplus[h,h-2,r]$\\ &$1$&$[h-1,h-1,r]\oplus[h,h,r]\oplus[h,h-1,r+1]\oplus[h,h-1,r-1]$ \\ \hline
$h+\fft32$&$\fft12$& $[h-1,h-1,r+1]\oplus[h-1,h-1,r-1]\oplus[h,h,r+1]\oplus[h,h,r-1]$\par${}\oplus[h,h-1,r]\oplus[h,h-1,r]\oplus[h-1,h-2,r]\oplus[h,h-1,r+2]$\par${}\oplus[h,h-1,r-2]\oplus[h,h-2,r+1]\oplus[h,h-2,r-1]$\\
&$\fft32$&$[h-1,h-1,r+1]\oplus[h-1,h-1,r-1]\oplus[h,h-1,r]$ \\ \hline
$h+2$&$0$&$[h-1,h-1,r]\oplus[h,h,r]\oplus[h-2,h-2,r]\oplus[h,h-1,r+1]\oplus[h,h-1,r-1]$\par${}\oplus[h-1,h-2,r+1]\oplus[h-1,h-2,r-1]\oplus[h,h-2,r]$\par${}\oplus[h,h-2,r+2]\oplus[h,h-2,r-2]$\\
&$1$&$[h-1,h-1,r]\oplus[h-1,h-1,r]\oplus[h-1,h-1,r+2]\oplus[h-1,h-1,r-2]$\par${}\oplus[h,h-1,r+1]\oplus[h,h-1,r-1]\oplus[h-1,h-2,r+1]$\par${}\oplus[h-1,h-2,r-1]\oplus[h,h-2,r]$\\
&$2$&$[h-1,h-1,r]$ \\ \hline
$h+\fft52$&$\fft12$&$[h-1,h-1,r+1]\oplus[h-1,h-1,r-1]\oplus[h-2,h-2,r+1]$\par${}\oplus[h-2,h-2,r-1]\oplus[h,h-1,r]\oplus[h-1,h-2,r]\oplus[h-1,h-2,r]$\par${}\oplus[h-1,h-2,r+2]\oplus[h-1,h-2,r-2]\oplus[h,h-2,r+1]\oplus[h,h-2,r-1]$\\
&$\fft32$&$[h-1,h-1,r+1]\oplus[h-1,h-1,r-1]\oplus[h-1,h-2,r]$\\ \hline
$h+3$&$0$&$[h-2,h-2,r]\oplus[h-2,h-2,r+2]\oplus[h-2,h-2,r-2]$\par${}\oplus[h-1,h-2,r+1]\oplus[h-1,h-2,r-1]\oplus[h,h-2,r]$\\
&$1$&$[h-1,h-1,r]\oplus[h-2,h-2,r]\oplus[h-1,h-2,r+1]\oplus[h-1,h-2,r-1]$\\ \hline 
$h+\fft72$&$\fft12$& $[h-2,h-2,r+1]\oplus[h-2,h-2,r-1]\oplus[h-1,h-2,r]$\\ \hline
$h+4$&$0$& $[h-2,h-2,r]$ \\ \hline
\end{tabular}

\caption{$\frac{1}{3}$-BPS multiplets  $\mathcal D_{\text{$\fft13$-BPS}}(h,0,h,h,r)$ of $osp(4|6)$, with $|r|<h$. Note there are special cases when $h-|r|<4$. For these cases, we must neglect the states that violate the condition $h_1\geq h_2\geq |h_3|$.}
\label{tbl:13BPS}

\end{table}

\section{One-loop free energy of supergravity on $\mathrm{AdS}_4\times S^7/\mathbb{Z}_k$}\label{sec:freeenergy}

With the Kaluza-Klein spectrum at hand, we may now turn to the computation of the one-loop free energy on global AdS$_4$.  Since supersymmetry is maintained level by level in the Kaluza-Klein spectrum, we organize the free energy as
\begin{equation}
F_{\text{1-loop}}=\sum_{n=0}^{\infty}F_{\text{1-loop},n},
\label{eq:Fsum}
\end{equation}
where $n$ is the Kaluza-Klein level.  The contribution at level $n$
can be written schematically in terms of a ratio of functional determinants:
\begin{equation}
Z_{\text{1-loop}, n}=\prod_{i\in K_n} \frac{\text{det}_F(-\nabla^2+c_i(E_i,s_i))^{\text{dim($s_i,E_i,n$)}}}{\text{det}_B(-\nabla^2+c_i(E_i,s_i))^{\text{dim($s_i,E_i,n$)}}},
\end{equation}
where $c_i(E,s)$ are functions of the spin and energy of the multiplets that are determined by the specific matter content, $\text{dim($s_i,E_i,n$)}$ is the dimension of the corresponding $su(4)$ representation of the multiplet and $K_n$ is the index set of supermultiplets at the $n$-th Kaluza-Klein level.

There are numerous methods for computing the functional determinants.  We use the spectral zeta function, which is defined as the Mellin transform of the trace of the heat kernel for the operator $-\nabla^2 +c_i$.  With $F_{\text{1-loop},n}=-\log Z_{\text{1-loop},n}$, one has \cite{hawking1977}
\begin{equation}
F_{\text{1-loop},n}=-\frac{1}{2}\sum_{i\in K_n}\text{dim($s_i,E_i,n$)}\zeta_{(E_i,s_i)}'(0)-\frac{\log L^2 \Lambda^2}{2}\sum_{i\in K_n}\text{dim($s_i,E_i,n$)}(\zeta_{E_i,s_i}(0)+\mu_i),
\end{equation}
where $L$ is the AdS$_4$ radius, $\Lambda$ is the mass cut off, $\zeta_{E_i,s_i}(z)$ is the spectral zeta function of the corresponding operator for the multiplet with energy $E_i$ and spin $s_i$, and $\mu_i$ is the zero mode contribution coming from possibly discrete eigenmodes for $-\nabla^2 +c_i$ with zero eigenvalue. However, for global AdS$_4$, the only case such an operator could possibly admit discrete eigenmodes is for harmonic two-forms \cite{CAMPORESI199457}, which only occur as generalized Grassmanian ghosts from the quantization of the three-form in the eleven dimensional supergravity action \cite{Bhattacharyya:2012ye}. Nevertheless, such ghosts are not included in the Kaluza-Klein spectrum (\ref{eq:susybranch}), and thus require a separate calculation. As the discrete spectrum has been accounted for in \cite{Bhattacharyya:2012ye}, in the following we shall focus on the continuous spectrum instead. 

The spectral zeta function for global AdS$_4$ with arbitrary spin and energy is known to be \cite{PhysRevD.47.3339} 
\begin{equation}
\zeta_{E_0,s}(z)=\mbox{Vol}(\text{AdS}_4) \frac{L^{2z-4}(2s+1)}{8\pi^2}\int_0^{\infty}d\lambda \frac{\lambda(\lambda^2+(s+\frac{1}{2})^2)\tanh(\pi(\lambda+i s))}{(\lambda^2+(E_0-\frac{3}{2})^2)^z},
\end{equation}
where the regularized volume of AdS$_4$ is $\frac{4\pi^2}{3}L^4$. The function $\zeta_{E_0,s}(z)$ can be analytically continued if one substitutes
\begin{equation}
\tanh(\pi(\lambda+is))=1-\frac{2}{1+e^{2\pi(\lambda+is)}},
\end{equation}
in which case it becomes a meromorphic function on $\mathbb{C}$ with simple poles at $z=1$ and $z=2$. In AdS$_4$, the spectral zeta function for arbitrary spin and energy at $z=0$ is given by
\begin{align}
    \zeta_{E,s}^B(0)&=\frac{2s+1}{24}\left[(E-\ft{3}{2})^4-(s+\ft{1}{2})^2\left(2(E-\ft{3}{2})^2+\ft{1}{6}\right)-\ft{7}{240}\right],\nn\\
    \zeta_{E,s}^F(0)&=-\frac{2s+1}{24}\left[(E-\ft{3}{2})^4-(s+\ft{1}{2})^2\left(2(E-\ft{3}{2})^2-\ft{1}{3}\right)+\ft{1}{30}\right],
\end{align}
for bosonic and fermionic fields, respectively. Note that the even-dimensional AdS case is distinct from the odd-dimensional one in that the bosonic and fermionic measures are different, thus giving rise to separate expressions for 
$\zeta_{E,s}^B$ and $\zeta_{E,s}^F$.

The derivative of the zeta function at $z=0$ is given by 
\begin{equation}
\begin{split}
    \zeta_{E_0,s}'(0)=&(-1)^{2\Delta}\Bigl[\frac{2s+1}{24}\left((2E_0s+E_0-3s-\ft{3}{2})^2-\ft{1}{6}(2E_0-3)^4\right)\\
    &+\frac{2s+1}{3}\zeta'(-3,E_0+\Delta)-(1+2s)(E_0-\ft{3}{2})\zeta'(-2,E_0+\Delta)\\
    &-\frac{2s+1}{6}(2s^2+2s-6E_0^2+18E_0-13)\zeta'(-1,E_0+\Delta)\\&
    -\frac{2s+1}{6}(2E_0-3)(E_0-s-2)(E_0+s-1)\zeta'(0,E_0+\Delta)\Bigr],
\end{split}
\end{equation}
where $\zeta'(s,a)=\partial\zeta(s,a)/\partial s$ is the derivative of the Hurwitz zeta function, and $\Delta=-1$ for bosons and $\Delta =-\frac{3}{2}$ for fermions. One might worry that a logarithmic divergence shows up in $\zeta'(0,E_0+\Delta)$ at $E_0=1$ for bosons or $E_0=\frac{3}{2}$ for fermions. However, as we have seen in section \ref{sec:KKspec}, the only boson with $E_0=1$ has spin zero, and therefore the factor $E_0+s-1$ vanishes, which suppresses the logarithmic divergence. In the fermionic case, the factor $2E_0-3$ plays a similar role. 

While the spectral zeta function regularizes the one-loop determinant in AdS$_4$, the sum over the KK tower, (\ref{eq:Fsum}), is divergent since
$\zeta_{E,s}(0)$ grows as $E^4$ and $\zeta_{E,s}'(0)$ grows as $E^4\log E$ for large $E$. Thus the KK sum must be regulated as well.  One possibility would be to attach some smooth factor $e^{-n}$, treating supergravity as effective only up to some energy scale, and therefore suppressing the contribution from high KK levels. A related approach is the introduction of a hard cutoff \cite{Beccaria:2014xda,Beccaria:2014qea}.  Alternatively, we follow the prescription of \cite{Mansfield:2002pa} and attach $z^n$ to each level, assuming $|z|<1$.  The regulated sum is then given by the finite term in the expansion as $z\to 1$. Note that the hard cutoff and $z^n$ regulators produce identical results for the case of polynomials in $n$.  In particular, polynomials in $n$ are regulated to zero in both cases.  The soft cutoff ($e^{-n}$ regulator), on the other hand, produces a different non-zero result.  In contrast with the odd-dimensional cases, however, log terms show up in AdS$_4$, and they make the hard cutoff prescription less convenient to implement.
The regulator $z^n$ was used in \cite{Ardehali:2013gra,Ardehali:2013xya} to calculate the one-loop free energy of supergravity on AdS$_5\times S^5/\mathbb{Z}_k$ and produced the correct holographic results. In the following we shall compute using the $z^n$ regulator, but will comment on using alternative regulators in Appendix~\ref{app:reg}.

Using the KK spectrum in Section \ref{sec:KKspec}, we find that the one-loop free energy of AdS$_4\times S^7/\mathbb{Z}_k$ with our choice of $z^n$ regulator can be expressed as the following sum:
\begin{align}\label{F1}
F_{\text{SUGRA}}^{(1)}&=\sum_{l=1}^{k-1}[c_1(l,0)\log(l)z^{2l-2}+c_2(l,0)\log(l+1)z^{2l-2}-2\zeta'(0,l)z^{2l-2}]\nn\\
&+\sum_{l=0}^{k-1}\sum_{m=1}^{\infty}\Bigl[c_1(l,m)\log(km/2+l)z^{km+2l-2}+c_2(l,m)\log(km/2+l+1)z^{km+2l-2}\nn\\
&\kern4.5em-2(1+m)\zeta'(0,km/2+l)z^{km+2l-2}\Bigr].
\end{align}
(Note that the $n=0$ Kaluza-Klein level includes ghost contributions for the massless fields.  However, the resulting expression for $F_{\text{1-loop, }n=0}$ fits the general pattern for $n>0$.)
The functions $c_1(l,m)$ and $c_2(l,m)$ are polynomials in $l$ and $m$, and their explicit forms are given in Appendix~\ref{app:c12}. As a consequence of $\mathcal{N}=6$ supersymmetry, the polynomial term is canceled completely, and the rest gives partial cancellations. Note the $\log L$ term disappears after summing $\zeta_{E,s}(0)$ over the KK tower. This is consistent with the analysis in \cite{Bhattacharyya:2012ye}, in which the $\log L$ term in the free energy only arises from the zero modes, which are contributions from the discrete spectrum.

The calculation of the regularized sum is somewhat lengthy, and in order to illustrate the general procedure, we consider the following sum:
\begin{equation}
S_1=\sum_{l=0}^{k-1}\sum_{m=1}^{\infty}c_1(l,m)\log(km+l)z^{2km+2l-2}.
\end{equation}
The $m$ sum can be rewritten in terms of the derivative $\partial/\partial s$ of the Hurwitz-Lerch function $\Phi(z,s,a)=\sum_{m=0}^{\infty}z^m(a+m)^{-s}$.  To do so, we expand $c_1(l,m)$ as a polynomial in $km+l$, so that it may be combined with the argument of the log. The treatment for the sum
\begin{equation}
S_2=-2\sum_{l=0}^{k-1}\sum_{m=1}^{\infty}(1+m)\zeta'(0,km/2+l)z^{km+2l-2},
\end{equation}
is somewhat different.  Using $\zeta'(0,x)=\log\Gamma(x)-\frac{1}{2}\log2\pi$ and $\log\Gamma(n+1)=\sum_{i=1}^{n}\log i$, this can be rewritten as a sum over logs, which can then be re-expressed in terms of the Hurwitz-Lerch function.

As a result, the regulated $F_{\text{SUGRA}}^{(1)}$ can be written in terms of the derivative of $\Phi(z,s,a)$ along with elementary functions. The Hurwitz-Lerch function can then be expanded in its first argument by
\begin{equation}\label{lerchexpansion}
\Phi(z,s,a)=z^{-a}\left(\Gamma(1-s)(-\log z)^{s-1}+\sum_{i=0}^{\infty}\zeta(s-i,a)\frac{\log^i(z)}{i!}\right),
\end{equation}
valid for $|\log z|<2\pi$, $s\notin\mathbb N^+$ and $-a\notin\mathbb N^0$. To obtain the regulated sum, we consider the expansion around $z\to 1^{-}$, and take the finite term. 
Recall that $k=1,2$ are special cases preserving $\mathcal{N}=8$ supersymmetry. For $k\geq 3$, we obtain the regulated expression
\begin{align}\label{exactanswer1}
   F_{\text{SUGRA}}^{(1)}=&\frac{40049}{72576 k}+\frac{1}{24}\left(11+6k-2k^2\right)\log\frac{k}{2}+\left(-\frac{35}{12 k}+\frac{k}{6}\right)\left(\log\Gamma({k}/{2})-\fft12\log2\pi\right)\nn\\
   &+\frac{k}{288\pi^2}\left(204+10k^2+k^4\right)\zeta(3)+\frac{17k^3}{48\pi^4}\zeta(5)-\frac{5k^2\zeta(7)}{64\pi^6}\nn\\
   &+\sum_{l=1}^{k-1}\biggl[\frac{k^5}{72}\zeta'\left(-6,\frac{2l}{k}\right)+\frac{17k^3}{36}\zeta'\left(-4,\frac{2l}{k}\right)-\frac{k^2}{36}(k-2l)(5l(k-l)-12)\zeta'\left(-3,\frac{2l}{k}\right)\nn\\
   &\kern3em+\frac{5k}{12}l(k-l)(l(k-l)+2)\zeta'\left(-2,\frac{2l}{k}\right)\nn\\
   &\kern3em-\frac{1}{36}(k-2l)\left(84+l(k-l)\left(3l^2-3kl-k^2-10\right)\right)\zeta'\left(-1,\frac{2l}{k}\right)\nn\\
   &\kern3em-\frac{2l(k-l)}{k}\zeta'\left(0,\frac{2l}{k}\right)\biggr],
\end{align}
for the one-loop free energy of 11-dimensional supergravity on AdS$_4\times S^7/\mathbb Z_k$. The discrete sum over $k$ naturally comes from the $\mathbb{Z}_k$ orbifolding of $S^7$.

In order to work out the sum over $k$, one needs Hurwitz zeta identities of the form
\begin{equation}\label{unknownidentity}
\sum_{l=0}^{k-1}l^m\zeta'(-n,\frac{l}{k}),
\end{equation}
generalizing the multiplication formula of the Hurwitz zeta function, which would be the case for $m=0$. However, whether such an identity has a closed form is, to our knowledge, unknown to the literature, and we have failed to find one. Nevertheless, the sum in (\ref{exactanswer1}) does have curious symmetries.  In particular, we observe that it can be rewritten into the following form:
\begin{equation}
\sum_{l=1}^{k-1}p_{i}(l)\Bigl(\zeta'(-n,{2l}/{k})+(-1)^{n}\zeta'(-n,2(1-{l}/{k}))\Bigr),
\end{equation}
where $p_{i}(l)$ are polynomial coefficients, although it is not apparent why one has such a symmetry. This form of the summand allows it to be written as a Clausen function using the relation
\begin{equation}
\zeta'(-n,x)+(-1)^n\zeta'(-n,1-x)=\frac{(-1)^{\lfloor\frac{n}{2}\rfloor}n!}{(2\pi)^n}\mathrm{Cl}_{n+1}(2\pi x),\qquad n\in \mathbb{N}^0,
\end{equation}
so that it may be turned into a sum of the form $\sum_{r=1}^{k-1}\left. \frac{d^m}{d^mz}\right|_{z={\pi r}/{k}}\cot^n(z) \zeta(a,{\pi r}/{k})$ with various $m, a, n\in \mathbb{Z}$. However, the closed form of such sums is not known, and it is not clear whether such a procedure would yield any additional physical insights.

In any case, for small values of $k$, the expressions in (\ref{exactanswer1}) can be simplified, and we list the result in Table~\ref{tbl:smallk}.  Note that the $k=1,2$ cases are computed separately, as they are special cases with $\mathcal{N}=8$ supersymmetry.  We see that the $k=3$ case has an extra transcendental part given in terms of the polygamma function, while the others have simple transcendental part. In fact, examination of additional small $k$ cases strongly suggests that $k=1,2,4$ are the only ones where $F_{\text{SUGRA}}^{(1)}$ has a simple transcendental part. It is curious that these values of $k$ are exactly those such that the $\mathbb Z_k$ action is in $SU(4)$, which is the $\mathcal{N}=6$ R-symmetry.

\begin{table}[t]
\centering
\setlength\extrarowheight{3pt}
    \begin{tabular}{ | l | p{8cm} | l |}
    \hline
    $k$ & $F_{\text{SUGRA}}^{(1)}$& Numerical Value  \\ \hline
    $1$ & $\frac{40049}{72576}+\frac{215\zeta(3)}{288\pi^2}+\frac{12\zeta(5)}{48\pi^4}-\frac{5\zeta(7)}{64\pi^6}-\frac{5\log 2}{4}$&$-0.220002$  \\[3pt] \hline
    $2$ & $\frac{40049}{145152}+\frac{215 \zeta (3)}{72 \pi ^2}+\frac{17 \zeta (5)}{3 \pi ^4}-\frac{5 \zeta (7)}{\pi ^6}$&$0.694679$ \\[3pt] \hline
    $3$ & $\frac{40049}{217728}+\frac{85 \zeta (3)}{32 \pi ^2}+\frac{17 \zeta (5)}{144 \pi ^4}-\frac{5 \zeta (7)}{192 \pi ^6}+\frac{\log2}{4}-\frac{\log3}{3}$\par
    ${}-\frac{\pi }{9 \sqrt{3}}+\frac{17 \psi_1({1}/{3})}{108 \sqrt{3} \pi }+\frac{\psi_3({1}/{3})}{432 \sqrt{3} \pi ^3} $\vspace*{3pt}&$ 0.427319$\\[3pt] \hline
    $4$& $\frac{40049}{290304}+\frac{265 \zeta (3)}{72 \pi ^2}+\frac{17 \zeta (5)}{6 \pi ^4}-\frac{5 \zeta (7)}{2 \pi ^6}-\frac{\log 2}{2}$& $0.267190$\\[3pt] \hline
    \end{tabular}
\caption{The holographic one-loop ABJM free energy for $k\le4$.}
\label{tbl:smallk}
\end{table}

\subsection{Asymptotic expansion of $F_{\text{SUGRA}}^{(1)}$ for large $k$}

The orbifold summation in the holographic result, (\ref{exactanswer1}), makes it a somewhat unwieldy expression.  However, this sum may be performed in the large-$k$ limit, allowing us to compare with the ABJM partition function in the corresponding limit.  Here it is important to note that, while taking $k\to\infty$ is generally considered the IIA limit, we can nevertheless remain in the M-theory limit by working with large but not infinite $k$, so long as we stay in the regime $k^5\ll N$ \cite{Aharony:2008ug}.  

\begin{figure}[t]
  \centering
  \includegraphics[width=4in]{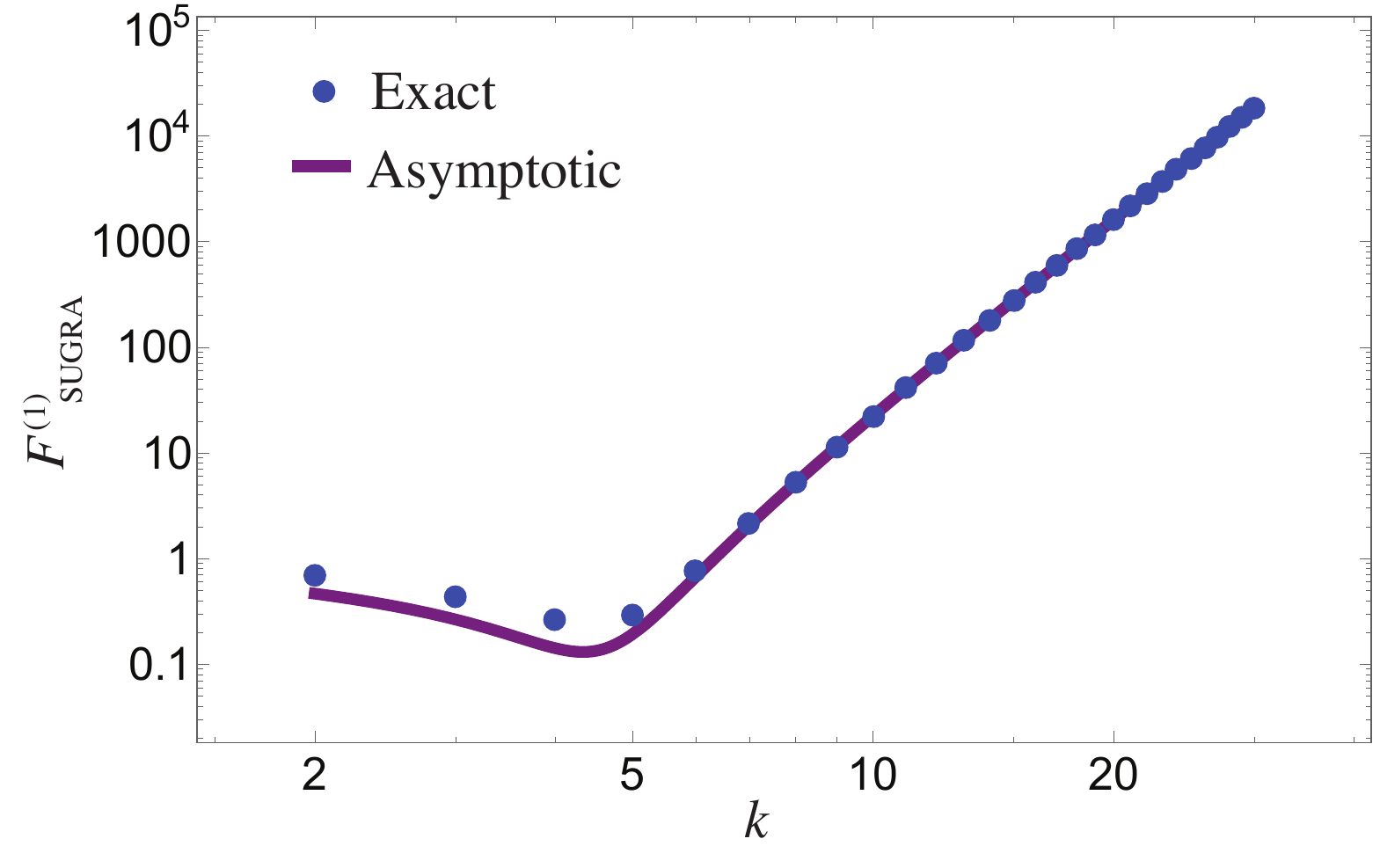}
  \caption{A log-log plot of the exact $F_{\text{SUGRA}}^{(1)}$ given in (\ref{exactanswer1}) along with its asymptotic expansion (\ref{Fasym}).}
   \label{theonlyfig}
\end{figure}

In the large-$k$ limit, we use the Euler-Maclaurin formula to rewrite the sum over $k$ in (\ref{exactanswer1}) according to:
\begin{align}
    \sum_{l=0}^{k}l^a\zeta(s,&\frac{2l}{k})=\nn\\
    &k^{a+1}\left(-\sum_{i=1}^a\frac{\Gamma(s-i)\Gamma(a+1)}{2^i \Gamma(s)\Gamma(a+2-i)}\zeta(s-i,2)+\frac{\Gamma(s-a-1)\Gamma(a+1)}{2^{a+1}\Gamma(s)}\right)+\frac{k^a}{2}\zeta(s,2)\nn\\
    &+\sum_{r=1}^p\frac{k^{a-2r+1}B_{2r}}{2r}\sum_{i=1}^{2r-1}\frac{\Gamma(a+1)\Gamma(s+2r-i-1)(-2)^{2r-1-i}}{\Gamma(i+1)\Gamma(a-i+1)\Gamma(2r-i)\Gamma(s)}\zeta(s+2r-i-1,2)\nn\\& +\mathcal{O}(k^{a-2p-1}),
\end{align}
where we take $a\geq 1$, $a-2p+1\geq 1$ and $s<0$.
The resulting expression for the one-loop free energy then becomes
\begin{equation}\label{Fasym}
F_{\text{SUGRA}}^{(1)}=\frac{25\zeta(7)}{1024\pi^6}k^6-\frac{3\zeta(5)}{128\pi^4}k^4-\frac{\zeta(3)}{18\pi^2}k^2-\frac{389}{945}\log\fft{k}2+C+\mathcal{O}\left(\frac{1}{k}\right),
\end{equation}
where
\begin{equation}
C=\frac{20 \log\mathcal{A}}{3}+\frac{5 \zeta'(-5)}{6}+\frac{13\zeta'(-3)}{3}-\frac{343}{3240}-\frac{556 \log 2\pi }{945},
\end{equation}
and $\mathcal{A}$ is the Glashier constant. We see that $F_{\text{SUGRA}}^{(1)}$ grows as $k^6$, with only even $k$ powers in the large $k$ expansion up to $\mathcal{O}({1}/{k})$. It is intriguing why this is the case, and also whether the coefficients of the asymptotic expansion have any physical meaning. For practical purposes, however, one can see the asymptotic expansion fits very well, even for moderate values of $k$, as can be seen in Fig.~\ref{theonlyfig}. 

The holographic computation of $F_{SUGRA}^{(1)}$ can now be compared with the ABJM result $F_{\text{ABJM}}^{(1)}$ given in (\ref{fabjm}).  Dropping the $\log L$ term, which is accounted for by the supergravity zero modes \cite{Bhattacharyya:2012ye}, we find
\begin{equation}\label{abjm}
F_{\text{ABJM},k}^{(1)}=-\fft12\log\fft{k\pi}8-A(k)= \frac{\zeta(3)}{8\pi^2}k^2-\frac{1}{3}\log\fft{\pi^2k}{32}-2\zeta'(-1)+\mathcal O\left(\fft1{k^2}\right).
\end{equation}
This asymptotic behavior is rather different from the holographic result, (\ref{Fasym}), as it grows as $k^2$ instead of $k^6$. Moreover, although $F_{\text{SUGRA}}^{(1)}$ has a sub-sub leading $k^2$ term, its coefficient does not match with the leading coefficient of $F_{\text{ABJM},k}^{(1)}$ either. 

\section{Discussion}
\label{sec:disc}

Our main conclusion is that, while the holographic computation of the ABJM free energy agrees with the matrix model result at leading $N^{3/2}$ order, there is disagreement at the one-loop (i.e.\ $N^0$) order.  In particular, we found that $F_{\text{SUGRA}}^{(1)}\sim k^6$,
while $F_{\text{ABJM}}^{(1)}\sim k^2$ in the M-theory limit.  It remains a puzzle as to how this discrepancy may be resolved.  However, it should be noted that there are several subtleties to the holographic calculation.  Before addressing some of these issues, we recall how it was performed.  While the full M-theory dual lives in 11 dimensions, we focused on its supergravity limit and immediately went down to four dimensions by Kaluza-Klein reducing on $S^7/\mathbb Z_k$.  We then regulated the one-loop determinants using the spectral zeta function on global AdS$_4$.  In contrast with a manifestly 11-dimensional calculation, we then had to introduce a second regularization when summing over the KK tower in order to address the short-distance divergences on the $S^7$ orbifold.

The regularization of the KK sum was performed by attaching a factor $z^n$ to the $n$-th KK level and then taking the finite part in the limit $z\to1^-$.
While this regularization scheme has been used successfully in the past
\cite{Mansfield:2002pa,Ardehali:2013gra,Ardehali:2013xya}, it has an undesirable feature in that a slight modification of the regulator from $z^n$ to $z^{an}$ with some constant $a$ will produce a finite shift, thus leading to a potential ambiguity in the regulated partition function.  Moreover, in some cases, such as the IIA limit, there is a log divergence, and not just power law divergences.  Removing the log divergence then leads to a further ambiguity from the constant pertaining to the log.  It is possible that an improved regulator will remove this apparent scheme dependence and lead to agreement with the matrix computation of $F_{\text{ABJM}}^{(1)}$. We have more to say about the ambiguity in the $z^n$ regularization method in Appendix~\ref{app:reg}.

Assuming the discrepancy between $F_{\text{SUGRA}}^{(1)}$ and $F_{\text{ABJM}}^{(1)}$ persists even when accounting for possible regularization scheme dependence, it then suggests that $F_{\text{M-theory}}^{(1)}\neq F_{\text{SUGRA}}^{(1)}$, assuming the $\text{AdS}_4/\text{CFT}_3$ correspondence holds at the quantum level.  What this would indicate is that M-theory has quantum behavior that is distinct from that of quantum supergravity.
(This is already evident at $\mathcal O(N^{1/2})$, which arises from an eight-derivative correction to the supergravity action.)  In particular, the additional M-theory contributions to the one-loop partition function would have to include terms $\sim k^6$ and $k^4$ with precisely the same coefficients so as to cancel the corresponding supergravity terms in (\ref{Fasym}) in order to have functional agreement with $F_{\text{ABJM}}^{(1)}$. Understanding how such terms might arise could shed light as to what effective one-loop terms one might consider in order to study quantum M-theory.

It is of course expected that the full M-theory spectrum would include additional towers of long multiplets of $Osp(4|6)$ (including higher-spin multiplets).  For theories dual to an odd-dimensional AdS bulk, the contribution of a long multiplet to the sphere partition function vanishes because of a complete cancellation between fermions and bosons.  However, this is no longer the case when working with an even-dimensional AdS bulk, as the boson and fermion measures are now different.  For example, the one-loop free energy for an $\mathcal N=6$ long representation $\mathcal D(E_0,s,0,0,0)$
is given by
\begin{align}
    &F^{(1)}_{\mathcal{N}=6\text{ Long}}=\nn\\
    &\quad\frac{8}{3}(16 s^3+24 s^2+50 s+21)\log(\Lambda^2L^2)-\frac{7}{2} (2 s+1) (5 s (s+1)+14) \log((E_0+1)(E_0+2)) \nn\\
    &\quad-\frac{1}{4} (2 s+1) (15 s (s+1)+28) \log (E_0 (E_0+3))-\frac{1}{12} s (s+1) (2 s+1) \log ((E_0-1) (E_0+4)),
\end{align}
for integer spin $s$.

It is reasonable to assume that the multiplicity, i.e.\ the $su(4)$ dimension formula, still gives $\mathcal{O}(E_0^5)$ dependence. The number of charges divisible by $k$ at a given KK level should be proportional to the KK level, and since one might also expect $E_0\sim n$, one may conclude that the multiplicity of such long representations is on the order of $\mathcal{O}(E_0^6)$. Using the $z^n$ regulator, one can directly see that $\sum E^6_0\log(E_0+l)\sim \zeta(7)k^6$, if one assumes that $E_0\sim n\sim k m$. with $m$ being summed over. This indicates that, with proper organization of the long multiplets, it may be possible to cancel the leading $k^6$ behavior seen in the sum over BPS states. However, the exact Kaluza-Klein spectrum of quantum M-theory beyond supergravity is still unknown, and even the task of enumerating the massive IIA spectrum in a curved background is difficult. It would be interesting to see whether the asymptotic behavior of the supergravity partition function, (\ref{Fasym}), gives reasonable constraints on the possible Ansat\"se for the spectrum of long multiplets in the full M-theory.

Another subtle issue is the quantum inequivalence in the on-shell treatment between classically equivalent matter contents. For example, massless two-forms contribute on-shell in the same manner as scalar fields, but they give a different result after quantization due to a topological contribution coming from ghosts \cite{DUFF1980179}. Nevertheless, such discrepancies are limited to the discrete part of the spectrum, and thus only enter into the $\log L$ coefficient. It therefore should not change the $k^6$ behavior found above.

Finally, it is interesting to consider the IIA-limit. Although our regularization method readily applies to the IIA limit, whose KK spectrum is simply the zero-charge sector of the full $\mathcal{N}=8$ theory, the IIA result computed in this way appears to be unphysical, as it carries a regulator-dependent ambiguity related to the removal of a divergent $\log(1-z)$ term in the limit $z\to 1^-$.  (See Appendix~\ref{app:reg} for details.)  It remains a rather puzzling question why such a $\log(1-z)$ term arises in the IIA-limit but not in the M-theory limit.

\acknowledgments

We would like to thank F.~Larsen and L.A.~Pando Zayas for useful discussions. WZ would like to thank D.~Speyer for helpful discussions on the branching $so(8)\to su(4)\oplus u(1)$, J.~Lagarias on finding the closed form of (\ref{unknownidentity}), and W.~Zhong for interesting discussions. This work was supported in part by the US Department of Energy under Grant No.~DE-SC0007859. The work of WZ was supported in part by the 2015 University of Michigan Mathematics Undergraduate Research program, and the 2016 University of Michigan Honors Summer Fellowship program.

\appendix

\section{The $q\equiv0$ mod $k$ states in the Kaluza-Klein spectrum}\label{app:specmodk}

Here we present the spectrum of 11-dimensional supergravity on $S^7/\mathbb Z_k$.  This is essentially the subset of the states on $S^7$ branched into $so(8)\to su(4)\oplus u(1)$ as shown in Table~\ref{tbl:Zkspect} that have $u(1)$ charges $q\equiv 0\mod k$.
Although the spectrum superficially appears different depending on even or odd $k$,
the resulting expressions can be written in general as, e.g., in (\ref{F1}).

The Kaluza-Klein spectrum is shown in Table~\ref{tbl:Zkq}.  For even orbifolds,
$[a,b,c|e]$ stands for
\begin{equation}
[a,b,c|e]=\sum\limits_{p=0}^{m}(2km+l-kp+a,b,l+kp+c),
\end{equation}
with $E_0=km+l+e$, for $0\leq l\leq k-2$. For $l=k-1$, the summation range of $p$ will be modified for some of the matter contents, and we mark it as $[a,b,c|e]_{*}$ and $[a,b,c|e]_{\dagger}$ for $\sum\limits_{p=-1}^{2m}$ and $\sum\limits_{p=0}^{2m+1}$ respectively.

For odd orbifolds, we have instead
\begin{equation}
[a,b,c|e]=\begin{cases}
\sum\limits_{p=0}^{m}((2k+1)(m-p)+l+a,b,l+(2k+1)p+c),&\!n=(2k+1)m+2l;\\
\sum\limits_{p=0}^{m-1}((2k+1)(m-p)+l+a,b,l+(2k+1)p+c+1),&\!n=(2k+1)m+2l+1.
\end{cases} 
\end{equation} 
In both cases, the lowest energy is $E_0={n}/{2}+e$. The structure of the odd orbifold special cases is more complicated. We use $[a,b,c|e]^{*}$ to denote extending the upper bound of the sum to $m$ and $[a,b,c|e]^{**}$ to denote extending the lower bound to $-1$ when $l=k-1$ and $n=(2k+1)m+2l+1$. We also use $[a,b,c|e]^{\dagger}$ to denote limiting the upper bound of the sum to $m-1$ and $[a,b,c|e]^{\dagger\dagger}$ for limiting the lower bound to 1 when $l=0$ and $n=(2k+1)m+2l$. Finally, we use $[a,b,c|e]^{-}$ and $[a,b,c|e]^{--}$ to denote the same change as the previous one but with $l=1$ and $n=(2k+1)m+2l$.

\begin{table}[t]
\centering
    \begin{tabular}{ | l | p{12.5cm}  |}
    \hline
    Spin& $su(4)\oplus u(1)$  \\ \hline
    $2$ & $[0,0,0|3]$ \\ \hline
    $\fft32$ & $[0,1,0|\frac{5}{2}]+[1,0,-1|\frac{5}{2}]_{\dagger}^{*}+[-1,0,1|\frac{5}{2}]_{*}^{**}+[1,0,-1|\frac{7}{2}]+[-1,1,-1|\frac{7}{2}]$\par${}+[-1,0,1|\frac{7}{2}]$\\ \hline
    $1$& $[0,0,0|2]+[1,0,1|2]+[1,1,-1|2]_{\dagger}^{*}+[-1,1,1|2]_{*}^{**}+[0,0,0|3]^{\dagger}+[0,0,0|3]^{\dagger\dagger}$\par${}+[1,1,-1|3]+[-1,1,1|3]+[0,1,-2|3]^{\dagger\dagger}+[-2,1,0|3]^{\dagger}+[-1,2,-1|3]$\par${}+[2,0,-2|3]_{\dagger}^{*}+[-2,0,2|3]_{\dagger}^{*}+[-1,0,-1|4]+[0,0,0|4]^{\dagger\dagger\dagger}+[0,1,-2|4]^{\dagger\dagger}$\par${}+[-2,1,0|4]^{\dagger}$ \\ \hline
    $\fft12$& $[2,0,0|\frac{3}{2}]_{\dagger}^{*}+[0,1,0|\frac{3}{2}]_{*}+[0,0,2|\frac{3}{2}]_{*}^{**}+[1,0,-1|\frac{5}{2}]+[-1,1,-1|\frac{5}{2}]$\par${}+[-1,0,1|\frac{5}{2}]+[2,0,0|\frac{5}{2}]^{\dagger\dagger}+[0,0,2|\frac{5}{2}]^{\dagger}+[0,1,0|\frac{5}{2}]^{\dagger}+[0,1,0|\frac{5}{2}]^{\dagger\dagger}$\par${}+[2,1,-2|\frac{5}{2}]_{*}^{\dagger\dagger**}+[-2,1,2|\frac{5}{2}]_{\dagger}^{\dagger*}+[0,2,-2|\frac{5}{2}]^{\dagger\dagger}+[-2,2,0|\frac{5}{2}]^{\dagger}$\par${}+[-1,0,1|\frac{7}{2}]^{-}+[1,0,-1|\frac{7}{2}]^{--}+[0,0,-2|\frac{7}{2}]+[-2,0,0|\frac{7}{2}]+[0,1,0|\frac{7}{2}]^{\dagger\dagger\dagger}$\par${}+[-1,1,-1|\frac{7}{2}]+[-1,1,-1|\frac{7}{2}]+[-3,1,1|\frac{7}{2}]^{\dagger-*}+[1,1,-3|\frac{7}{2}]^{\dagger\dagger--**}$\par${}+[-2,2,0|\frac{7}{2}]^{\dagger}+[0,2,-2|\frac{7}{2}]^{\dagger\dagger}+[-1,1,-1|\frac{9}{2}]^{\dagger}+[0,0,-2|\frac{9}{2}]^{\dagger\dagger}+[-2,0,0|\frac{9}{2}]$ \\ \hline
    $0$ & $[1,0,1|1]_{*\dagger}^{***}+[1,0,1|2]+[3,0,-1|2]_{\dagger}^{*}+[-1,0,3|2]_{*}^{*}+[-1,2,-1|2]$\par${}+[1,1,-1|2]+[-1,1,1|2]+[-1,1,1|3]^{-}+[1,1,-1|3]^{--}+[1,2,-3|3]_{\dagger}^{\dagger\dagger--*}$\par${}+[-3,2,-1|3]_{*}^{\dagger-**}+[0,0,0|3]^{\dagger\dagger\dagger}+[-1,2,-1|3]+[0,1,-2|3]^{\dagger\dagger}$\par${}+[-2,1,0|3]^{\dagger}+[-1,0,-1|3]^{\dagger\dagger\dagger}+[1,0,1|3]^{\dagger\dagger\dagger}+[-1,0,-1|4]^{\dagger\dagger\dagger}$\par${}+[1,0,-3|4]_{\dagger}^{\dagger\dagger--*}+[-3,0,1|4]_{*}^{\dagger-**}+[-1,2,-1|4]^{\dagger\dagger\dagger}+[0,1,-2|4]^{\dagger\dagger}$\par${}+[-2,1,0|4]^{\dagger}+[-1,0,-1|5]^{\dagger\dagger\dagger}$ \\ \hline
    \end{tabular}
    \caption{The Kaluza-Klein spectrum of 11-dimensional supergravity on AdS$_4\times S^7/\mathbb Z_k$.  The $[a,b,c|e]$ notation is explained in the text.}
    \label{tbl:Zkq}
\end{table}

\section{Regulator dependence of the one-loop free energy}\label{app:reg}

In our calculation of $F_{\text{SUGRA}}^{(1)}$, we choose to attach $z^n$ with $z\in(0,1)$, and take the finite part of  $\lim_{z\to1^{-}}\sum_{n=0}^{\infty}F_{\text{SUGRA},n}^{(1)}z^{n}$. However, it is not \emph{a priori} clear why one should attach $z^n$ instead of, e.g. $z^{2n}$ or some other power. In general, different choices of the regulator produce a finite shift in the partition function, and thus make the regulated result ambiguous.  Consider, for example, the regulated sum
\begin{align}
S_1&=\sum_{a,b}\sum_{l=0}^{k-1}\sum_{m=0}^{\infty}c_{1ab}m^b\left(1+\frac{l}{k}\right)^a\log(km+k+l)z^{2km+2l} z^{2k-2}\nn\\
&=-\frac{\partial}{\partial s}\left[\sum_{ab}c_{1ab}z^{2k-2}k^{-s}\sum_{j=0}^b\binom{b}{j}(-1)^{b-j}\sum_{l=0}^{k-1}z^{2l}\left(1+\frac{l}{k}\right)^{a+b-j}\!\!\Phi\left(z^{2k},s-j,1+\frac{l}{k}\right)\right]_{s=0}\!\!,
\end{align}
which is similar to the case we face for the two logarithmic sums in (\ref{F1}).  Expanding the Hurwitz-Lerch function using (\ref{lerchexpansion}) gives
\begin{align}\label{S1Z}
    S_1=\sum_{a,b}c_{1ab}z^{-2}\sum_{j=0}^b(-1)^{b-j}\binom{b}{j}\sum_{l=0}^{k-1}\left(1+\frac{l}{k}\right)^{a+b-j}\biggl[\frac{j!(H_j-\gamma-\log(-2\log z))}{(-2k\log z)^{j+1}}\nn\\+\sum_{i=0}^{\infty}\left(\log k\, \zeta\left(-i-j,1+\frac{l}{k}\right)-\zeta'\left(-i-j,1+\frac{l}{k}\right)\right)\frac{(2k\log z)^i}{i!}\biggr],
\end{align}
Here one observes that replacing the $z^n$ regulator by $z^{an}$ (for some constant $a$) only changes the superficially divergent part, i.e. the first term in the inner sum of (\ref{S1Z}), through the $\log\log z$ factor. In the limit $z\to 1^{-}$, the superficially divergent term gives rise to $\log (1-z)$ terms in the $\mathcal{O}(1)$ part of the expansion, which makes the limit $z\to 1^{-}$ ill-defined in general. However, after summing over all terms in (\ref{F1}), we find that the $\log(1-z)$ terms in fact cancel with each other, and this makes our regularization procedure well defined.

In general, one might ask which regulator in the family of regulators $z^{a n}$ with arbitrary $a$ can produce a well-defined regularization procedure. The answer, to our surprise, is highly restrictive, namely ${1}/{a}$ has to be a divisor of 6. To see this, one can expand (\ref{F1}) using instead a $z^{a n}$ regulator.  We find that the resulting $\log(1-z)$ term has a coefficient 
\begin{equation}
\frac{(a-1) (2 a-1) (3 a-1) (6 a-1) \left(41 a^2+12 a+1\right)}{1008 a^6 k},
\end{equation}
which is zero for ${1}/{a}=1,2,3,6$. It would be interesting to further elucidate the relation between compatible regulators and the supersymmetry of the theory. Moreover, one consider regulating $F_{\text{SUGRA}}^{(1)}$ using a completely different procedure. Each different choice would give the regulated $F_{\text{SUGRA}}^{(1)}$ a finite shift. However, the finite shifts have asymptotic behaviors of at most the order $\mathcal{O}(k\log k)$. Therefore, while the regulated answer for finite $k$ is scheme dependent, the finite shift is not arbitrary. Moreover, the asymptotic expansion (\ref{Fasym}), is regulator independent up to the $k^2$ term. It is thus perhaps more intriguing to ask whether the coefficients of $k^6$, $k^4$ and $k^2$ in (\ref{Fasym}) have physical meaning as they appear to be fully scheme independent.

Finally we comment on the IIA limit. To reduce to the IIA spectrum on $\text{AdS}_4\times \mathbb{CP}^3$, we simply take the zero $u(1)$ charge sector. The sum of the zero charge sector using the $z^n$ regulator is given by
\begin{equation}
F_{\text{IIA}}^{(1)}=\frac{121}{864}+\frac{389}{945}(\gamma+\log (1-z))+\frac{20\log\mathcal{A}}{3}-\log 2\pi+\frac{5}{6}\zeta'(-5)+\frac{13}{3}\zeta'(-3),
\end{equation}
after taking the $\mathcal{O}(1)$ term in the $z\to 1^{-}$ expansion.
Curiously, we see a non-vanishing $\log(1-z)$ term, which makes the $z\to 1^{-}$ procedure ambiguous. Here one may consider using instead a $z^{an}$ regulator, with $a$ to be determined.  However, the regulator dependent versus independent terms are harder to disentangle, as we no longer have a parameter $k$ to play with.

Note that the IIA spectrum is exactly $\sum_{E_0} \mathcal D_{\frac{1}{3}\text{BPS}}(E_0,0,E_0,E_0,0)$ in terms of the $\mathcal{N}=6$ BPS multiplets, and the failure of the cancellation of the $\log(1-z)$ term for the IIA spectrum suggests that regulator compatibility also depends on how the different representations $\mathcal D_{\frac{1}{3}\text{BPS}}(E_0,0,E_0,E_0,r)$ fit into the KK spectrum.

\section{The polynomials $c_1(l,m)$ and $c_2(l,m)$}\label{app:c12}

The polynomials in $l$ and $m$ occurring in (\ref{F1}) are given by the following: 
\begin{align}\label{c1}
    c_1(l,m)=-\frac{1}{720} (m+1)\Bigl[&5 k^5 m^2\left(m^3-m^2+m-1\right)+9 k^4 m\left(6 m^3-6 m^2+m-1\right)\nn\\
    &+10 k^3 m^2 (17 m-5)-75 k^2 (m-1) m-840 k m+720\nn\\
     &+10 l\Bigl(k^4 m \left(6 m^3-6 m^2+m-1\right)+27 k^3 m^2 (2 m-1)\nn\\
     &\kern3.6em+2 k^2 m (56 m-5)-45 k m-168\Bigr)\nn\\
    &+30 l^2\Bigl(5 k^3 m^2 (2 m-1)+9 k^2 m (7 m-1)+78 k m-15\Bigr)\nn\\ 
    &+20 l^3\Bigl(5 k^2 m (7 m-1)+135 k m+78\Bigr)
    +150 l^4\Bigl(5 k m+9\Bigr)+300 l^5\Bigr],
\end{align}
\begin{align}\label{c2}
    c_2(l,m)=-\frac{1}{720} (m+1)\Bigl[&k^2 (m-1) m(5 k^3(m^3+m)+k^2(6 m^2+1)-70 k m+5)\nn\\
    &+10l k m\Bigl(k^3 (6 m^3-6 m^2+m-1)+3 k^2 m (2 m-1)\nn\\
    &\kern5em+k (14-56 m)+3\Bigr)\nn\\
    &+30 l^2\Bigl(5 k^3 m^2 (2 m-1)+k^2 m (7 m-1)-42 k m+1\Bigr)\nn\\
    &+20 l^3\Bigl(5 k^2 m (7 m-1)+15 k m-42\Bigr)
   +150 l^4 (5 k m+1)+300 l^5\Bigr].
\end{align}
Note that we do not observe the same $l\to k-l$ symmetry as in (\ref{exactanswer1}) directly in (\ref{c1}) and (\ref{c2}).

\bibliographystyle{JHEP}
\bibliography{reference}

\end{document}